\documentclass{article}

\usepackage{arxiv}

\usepackage[utf8]{inputenc} 
\usepackage[T1]{fontenc}    
\usepackage{hyperref}       
\usepackage{url}            
\usepackage{booktabs}       
\usepackage{amsfonts}       
\usepackage{nicefrac}       
\usepackage{microtype}      
\usepackage{cleveref}       

\usepackage{graphicx}
\usepackage{doi}
\usepackage{algorithm}
\usepackage{algorithmic}
\usepackage{float}
\title{Implementation of polygon guarding algorithms for art gallery problems}

\author{ Shiva Maleki\\
	Department of Mathematics and Computer
	Science\\
	Amirkabir University of Technology\\
	\texttt{shivamaleki112@aut.ac.ir} \\
	\And
	Ali Mohades \\
	Department of Mathematics and Computer
	Science\\
	Amirkabir University of Technology\\
	\texttt{mohades@aut.ac.ir} \\
}

\renewcommand{\shorttitle}{Shiva Maleki et al.}

\hypersetup{
pdftitle={Implementation of polygon guarding algorithms for art gallery problems},
pdfsubject={q-bio.NC, q-bio.QM},
pdfauthor={Shiva Maleki, Ali Mohades},
pdfkeywords={First keyword, Second keyword, More},
}

\begin{document}
\maketitle

\begin{abstract}
		Victor Klee introduce the art gallery problem during a conference in Stanford in August 1976 with that question: "How many guards are required to guard an art gallery?" In 1987, Ghosh provided an approximation algorithm for vertex guards problem \cite{ghosh2010approximation} that achieved $ O(\log n) $ approximation ratio.
	In 2017, Bhattacharya et al. \cite{bhattacharya2017approximability} presented a 6-approximation algorithm for guarding weak visibility polygons. In our paper, we first implement these algorithms and then we test them for different types of polygons. We compare their performance in terms of number of guards used by them. In the last part, we have provided a new algorithm that uses Ghosh's idea. Experiments show that this algorithm assigns near optimal guards for guarding the input polygons.
\end{abstract}


\section{Introduction}
Victor Klee introduce the art gallery problem during a conference in Stanford in August 1976  with this question: "How many guards are required to guard an art gallery?" We describe an art gallery as a simple polygon $ P $ with total of $ n $ vertices. A guard can be viewed as a point in $ P $. We say a point $ z \in P $ is visible from a guard $ g $ if the line segment $ gz $ lies inside $ P $ and dose not intersect the exterior of $ P $. If the guards are allowed to be placed just on vertices, we called vertex guards. If there is no restriction, guards are called point guards.
A polygon $ P $ is called weak visibility polygon if every point in $ P $ is visible from some point of  edge \cite{ghosh2007visibility}.

After Victor Klee posed the art gallery problem, V. Chv´atal established in \cite{chvatal1975combinatorial} that for simple polygon $ P $, 
$ \lfloor \frac{n}{3} \rfloor $ guards are always sufficient for guarding $ P $. If all edge of simple polygon $ P $ are vertical or horizontal, $ P $ is called simple orthogonal polygon. Kahn et al. \cite{kahn1983traditional} and O’Rourke \cite{o1983alternate} proved that simple orthogonal polygon $ P $ needs at most 
$ \lfloor \frac{n}{4} \rfloor $ guards. 

Lee and Lin 
\cite{lee1986computational} showed that the problem of computing a minimum number of guards for guarding a polygon is NP-Hard. In 2010, Ghosh \cite{ghosh2010approximation} presented an approximation algorithm for minimum vertex guard problem for simple polygons. The pseudo code of algorithm is:

\begin{algorithm}[H]
	\caption{An $ O(n^{4}) $-algorithm for computing a guard set $ S $ for all vertices of polygon $ P $}
	\begin{algorithmic} 
		\STATE Draw lines through every pair of vertices of $ P $
	and compute all convex components $ c_{1}, c_{2}, ..., c_{m} $ of $ P $
		\STATE Let  $C = (c_{1}, c_{2}, ..., c_{m}) $, $ N = (1, 2, ..., n) $  and  $ Q = \emptyset $
		\FOR {$ j=1 $ to $j=n $}
		
		\STATE construct the set $ F_{j} $ by adding those convex components of $ P $ that are totally visible from the vertex $ V_{j} $.
		\ENDFOR
		\WHILE {$ \mid C \mid \neq \emptyset $}
		\FOR {$ j \in N $}
		\STATE Find $ i \in N $ such that
		$\mid F_{j} \mid \leq \mid F_{i} \mid  $ and {$ i \neq j $}
		\ENDFOR
		\FOR{$ j \in N $}
		\STATE $ F_{j} := F_{j} -  F_{i} $
		\STATE $ C := C - F_{i} $
		\ENDFOR
		\ENDWHILE
	\end{algorithmic}
\end{algorithm}

In 2017, Bhattacharya et al. \cite{bhattacharya2017approximability} established a 6-approximation algorithm for vertex guarding a weak visibility polygon $ P $ that contains no holes and this algorithm has running time $ O(n^{2}) $.
Its pseudo code is:
\begin{algorithm}[H]
	\caption{An $ O(n^{2}) $-algorithm for computing a guard set $ S $ for all vertices of $ P $}
	\begin{algorithmic} 
		\STATE Compute $ SPT(u) $ and $ SPT(v) $
		\STATE Initialize all the vertices of $ P $ as unmarked
		\STATE Initialize $ B \leftarrow \emptyset $, $ S_{B} \leftarrow \emptyset $ and $ z \leftarrow u$
		\WHILE {there exists an unmarked vertex in $ P $ }
		\STATE $ z \leftarrow$ the first unmarked vertex on $ bd_{c}(u,v) $ in clockwise order from $ z $
		\IF{every unmarked vertex of $ bd_{c}(z, p_{v}(z)) $ is visible from $ p_{u}(z) $ or $ p_{v}(z) $}
		\STATE $ B \leftarrow B \cup \left\{z\right\}$  and 
		$ S_{B} \leftarrow S_{B} \cup \left\{p_{u}(z), p_{v}(z)\right\} $ 
		\STATE Mark all vertices of $ P $ that become visible from $ p_{u}(z) $ or $ p_{v}(z) $
		\STATE $ z \leftarrow p_{v}(z)  $
		\ELSE
		\STATE  $ z \prime \leftarrow $the first unmarked vertex on $ bd_{c}(z, v) $ in clockwise order
		\WHILE {every unmarked vertex of $ bd_{c}(p_{u}(z\prime), z\prime) $ is visible from $ p_{u}(z\prime) $ or $ p_{v}(z\prime) $}
		\STATE $ z \leftarrow z \prime $ and $ z \prime \leftarrow $ the first unmarked vertex on  $ bd_{c}(z \prime, v) $ in clockwise order 
		\ENDWHILE
		\STATE $ B \leftarrow B \cup \left\{z\right\} $ and $ S_{B} \leftarrow S_{B} \cup \left\{p_{u}(z), p_{v}(z)\right\}$
		\WHILE{there exists an unmarked vertex on $ bd_{c}(u, z) $}
		\STATE  $ w \leftarrow $ the first unmarked vertex on $ bd_{cc}(z, u) $ in counterclockwise order
		\STATE $ B \prime \leftarrow B \prime \cup \left\{w\right\} $ and $ S\prime_{B} \leftarrow S\prime_{B} \cup \left\{p_{u}(w), p_{v}(w)\right\} $
		\STATE Mark all vertices of $ P $ that become visible from $ p_{u}(w) $ or 
		$ p_{v}(w) $
		
		\ENDWHILE
		\ENDIF
		\ENDWHILE
		\STATE Reinitialize all the vertices of $ P $ that are visible from some guard in $ S_{B} $ as unmarked 
		\FOR {each vertex $ z \in B\prime $ chosen in reverse order of inclusion}
		\STATE Locate and mark each unmarked vertex visible from $ p_{v}(z) $ or $ p_{u}(z) $
		\IF {no new vertices get marked due to guards at $ p_{v}(z) $ or $ p_{u}(z) $}
		\STATE $ B\prime \leftarrow B\prime \setminus \left\{z\right\} $ and $ S\prime_{B} \leftarrow S\prime_{B} \setminus \left\{p_{u}(w), p_{v}(w)\right\} $
		\ENDIF
		\ENDFOR
		\STATE $ B \leftarrow B \cup B\prime $
		\RETURN the guard set $ S = S_{B} $
	\end{algorithmic}
\end{algorithm}

\subsection{Outline}
We implement the above mentioned algorithms and test both of them on weak visibility polygons. These weak visibility polygons are generated by a procedure presented in Section 2. Further, our new algorithm is tested on simple polygons which are generated by another procedure as mentioned in Section 3. We show experimentally that this algorithm assigns near optimal guards for guarding the input polygons.

\section{Test on weak visibility polygons}

\subsection{Algorithm for generate arbitrary weak visibility polygon }

We introduce an algorithm for generating arbitrary weak visibility polygons.

Let $ p = (k, 0) $ and $ q = (-k, 0) $:

Step 1: 
Choose $ n $ random points
$ x_{1}, x_{2}, ... , x_{n} $ 
in $ pq $ and sort them such that 

$ \mid px_{i} \mid > \mid px_{i+1} \mid $ for 
$ i \in N $.

Step 2: 
Choose $ n $ random angles 
$ \alpha_{1}, \alpha_{2}, ... , \alpha_{n}, $ 
in 
$ (0, \pi) $ and sort them such that

$ \mid \alpha_{j} \mid > \mid \alpha_{j+1} \mid $ for 
$ j \in N $. 

Step 3: Choose $ n $ arbitrary positive numbers 
$ r_{1}, r_{2}, ... , r_{n} $.

Step 4: For every $ i \in N $ compute $ \overrightarrow{y_{i}} $ as
$ \overrightarrow{y_{i}} = \overrightarrow{x_{i}} + (-r_{i}\cos \alpha_{i}, -r_{i}\sin \alpha_{i}) $.

Step 5: It is obvious that for every $ i \in N $, a quadrilateral with vertices $ x_{i}y_{i}y_{i+1}x_{i+1} $ is a convex quadrilateral so any point in this quadrilateral is visible from vertex $ x_{i} $ and $ y_{i} $. Choose four positive arbitrary numbers like 
$ w_{1}, w_{2}, w_{3}, w_{4} $ 
and compute 
$ z_{i} = \frac{w_{1}x_{i} + w_{2}y_{i} + w_{3}y_{i+1} + w_{4}x_{i+1}}{w_{1} + w_{2} + w_{3} + w_{4}} $. $ z_{i}  $ is a point in a quadrilateral with vertices $ x_{i}y_{i}y_{i+1}x_{i+1} $.

\begin{figure}[H]
	\centering
	\includegraphics[scale=0.4]{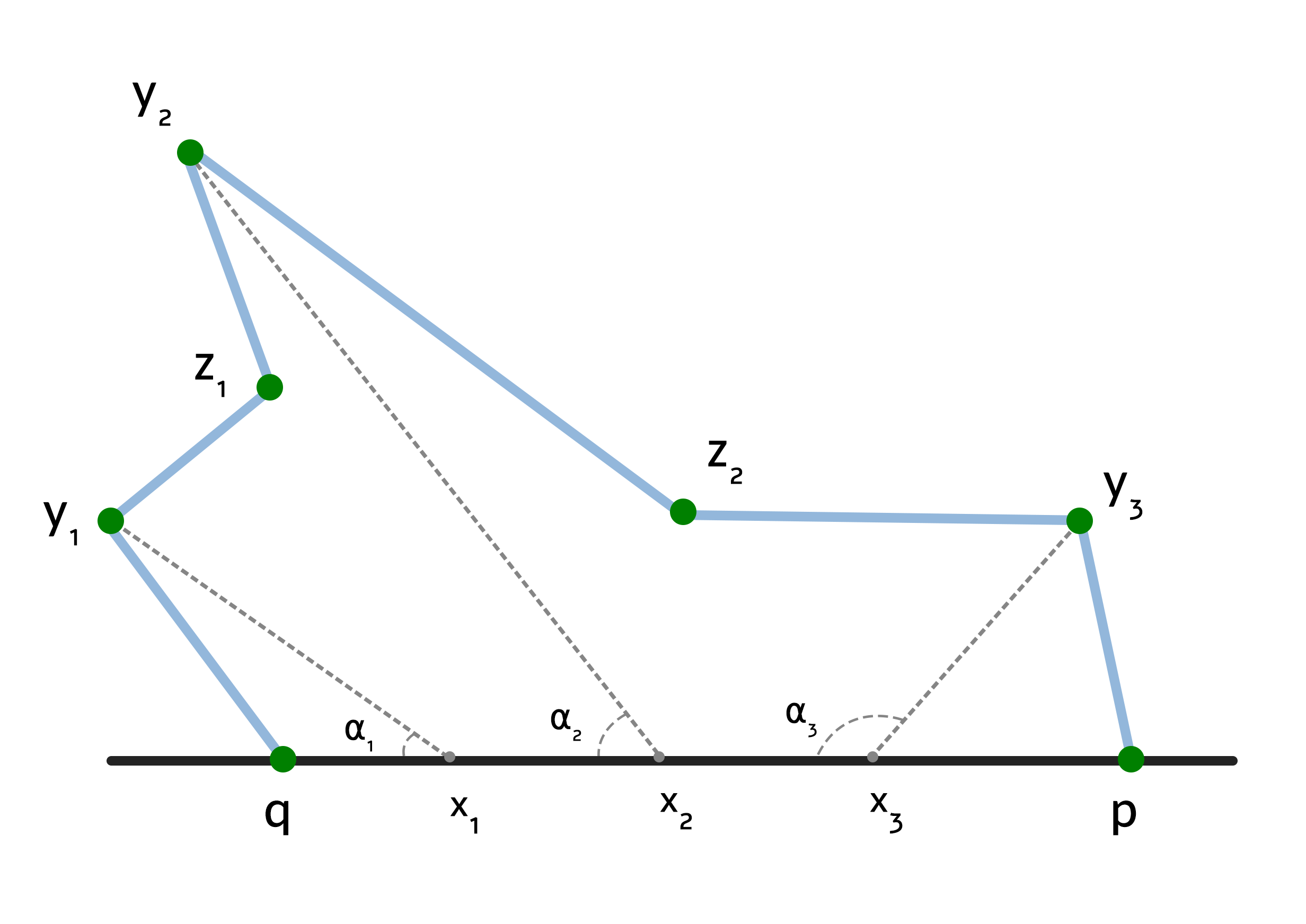}
	\caption{Generate weak visibility polygon with $ n = 3 $.}
	\label{fig:weak}
\end{figure}

It can be seen that a polygon with vertices 
$ qy_{1}z_{1}y_{2}z_{2}y_{3}z_{3} ... y_{n-1}z_{n-1}y_{n}p $
is a weak visibility polygon. (See figure \ref{fig:weak})

\subsection{Test on small weak visibility polygons with low value of reflex vertices}
The inputs of our test are the weak visibility polygons with number of vertices $ n $ are between 10 to 15 ($ 10 \leq n \leq 15  $) and number of reflex vertices $ r $ are between 2 to 3 ($ 2\leq r \leq 3 $). 

\begin{figure}[H]
	\centering
	\includegraphics[scale=0.6]{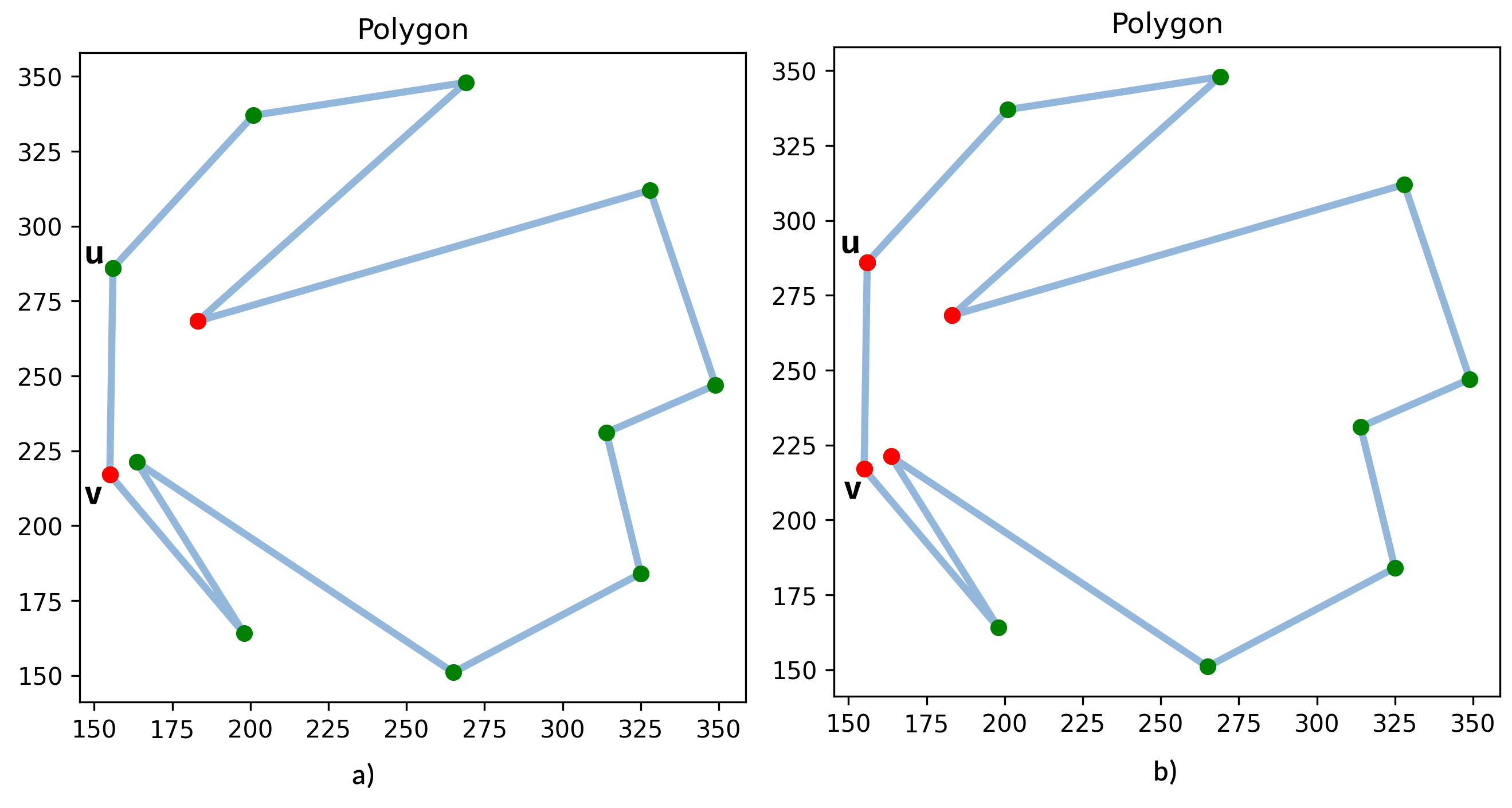}
	\caption{(a) For polygon with $ 12 $ vertices such that $ r = 3 $ guard set in Algorithm 1 is [(183.22, 268.31), (155.0, 217.0)] and it runs in 2.909 s. (b) guard set in Algorithm 2 is [(156.0, 286.0), (155.0, 217.0), (183.22, 268.31), (163.60, 221.22)] and it runs in 0.045 s. }
	\label{fig:test1_1}
\end{figure}

\begin{figure}[H]
	\centering
	\includegraphics[scale=0.6]{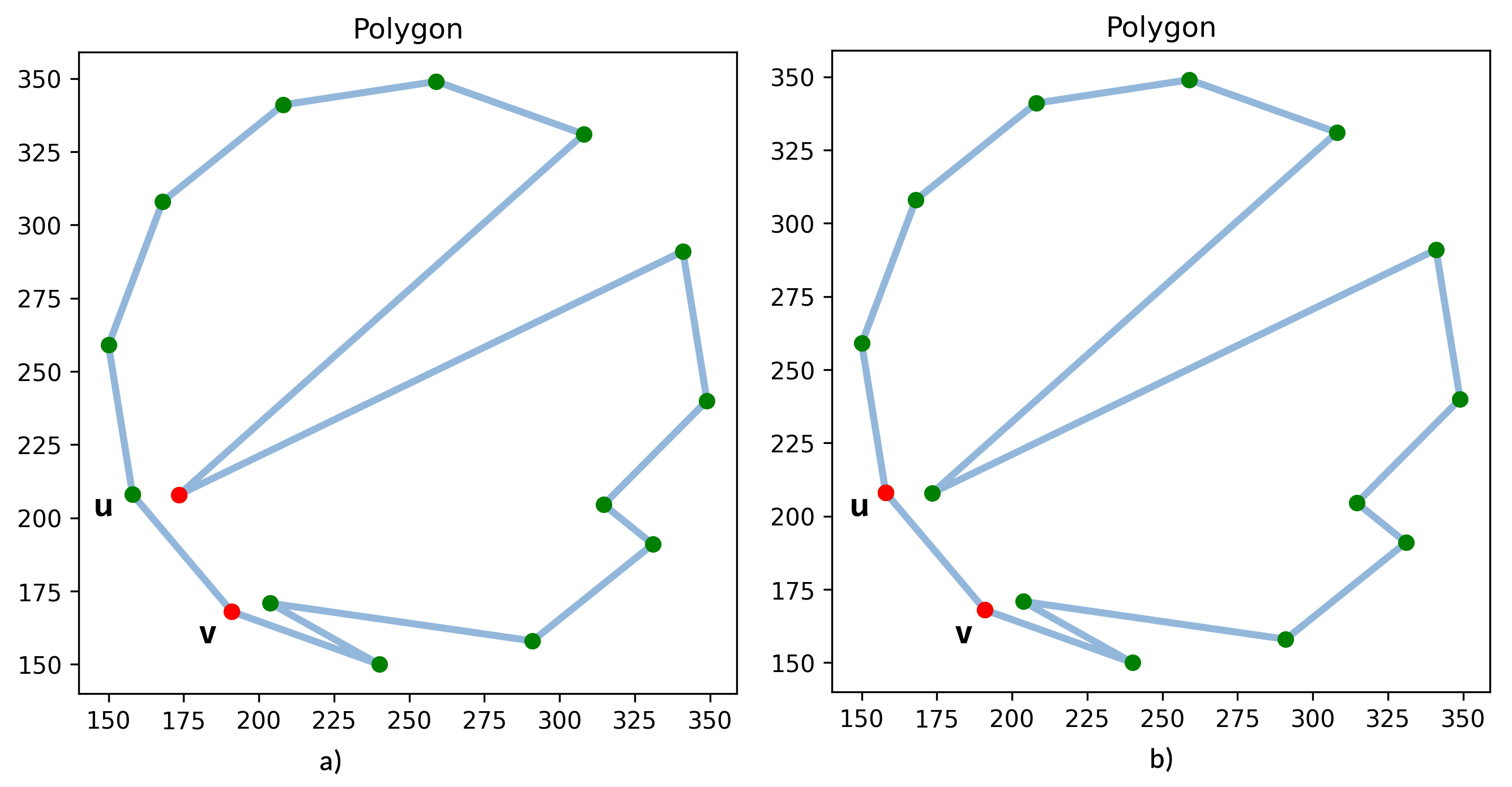}
	\caption{(a) For polygon with $ 15 $ vertices such that $ r = 3 $ guard set in Algorithm 1 is [(173.44, 207.91), (191.0, 168.0)] and it runs in 24.058 s. (b) Guard set in Algorithm 2 is [(158.0, 208.0), (191.0, 168.0)] and it runs in 0.0392 s. }
	\label{fig:test1_2}
\end{figure}

The outputs of test suggest that (see Figure~\ref{fig:test1_1} and Figure~\ref{fig:test1_2}) for a low value of $ n $ and $ r $, it is better to use Algorithm 1 for minimizing the number of vertex guards as Algorithm 2 uses more guards than Algorithm 1. Since Algorithm 2 is a constant approximation algorithm, Algorithm 1 performs like a constant time approximation algorithm for small values $ n $ and $ r $ experimentally.

Since the criteria of minimization is the number of guards rather than the running time which is an one time affair unlike online algorithms, Algorithm 1 is preferable even for weak visibility polygons.

\subsection{Test on small weak visibility polygons with number of reflex vertices $ r $ are roughly same and close to number of convex vertices}
In this test our inputs are also weak visibility polygons with the number of vertices $ n $ are between 10 to 31 ($ 10 \leq n \leq 31  $) and the number of reflex vertices $ r $ are roughly same and close to number of convex vertices $ (n-r) $.

\begin{figure}[H]
	\centering
	\includegraphics[scale=0.6]{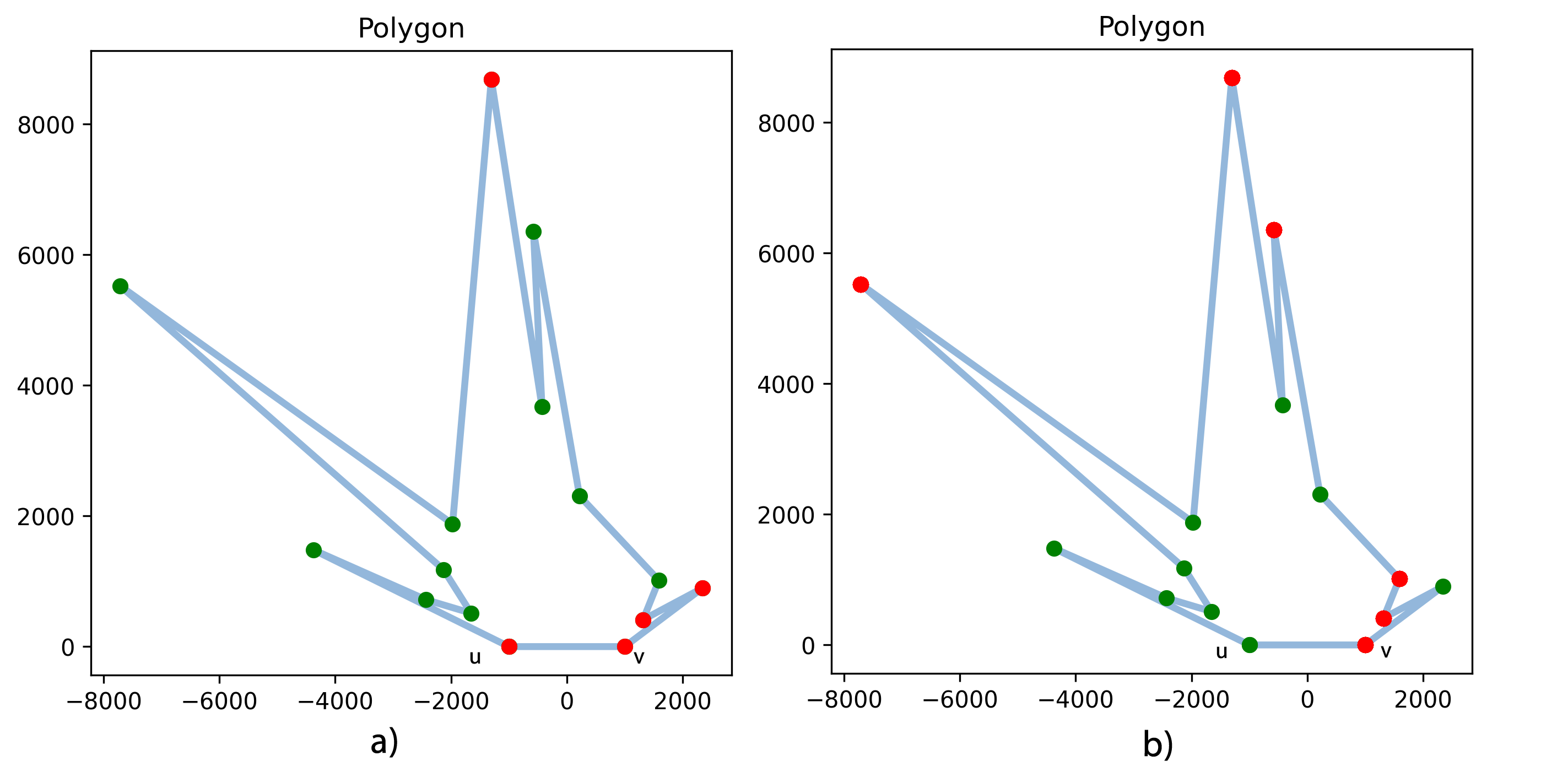}
	\caption{(a) For polygon with $ 15 $ vertices such that $ r = 7 $ size of guard set in Algorithm 1 is $ 5 $ and it runs in 0.2389 s. (b) Size of guard set in Algorithm 2 is $ 6 $ and it runs in 0.0718321 s. }
	\label{fig:test2_1}
\end{figure}

\begin{figure}[H]
	\centering
	\includegraphics[scale=0.6]{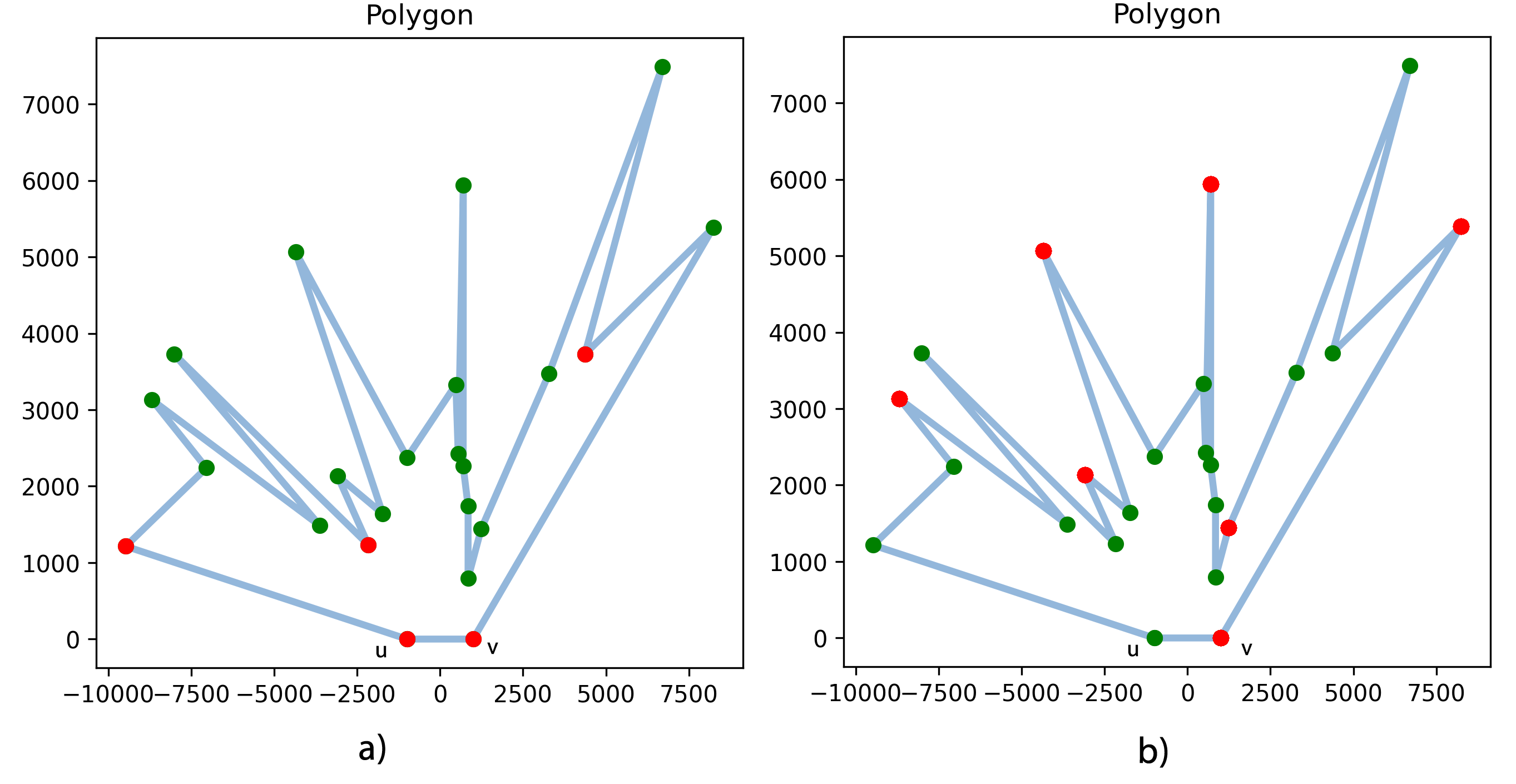}
	\caption{(a) For polygon with $ 23 $ vertices such that $ r = 13 $ size of guard set in Algorithm 1 is $ 3 $ and it runs in 0.373031 s. (b) Size of guard set in Algorithm 2 is $ 9 $ and it runs in 0.27948 s. }
	\label{fig:test2_2}
\end{figure}

\begin{figure}[H]
	\centering
	\includegraphics[scale=0.6]{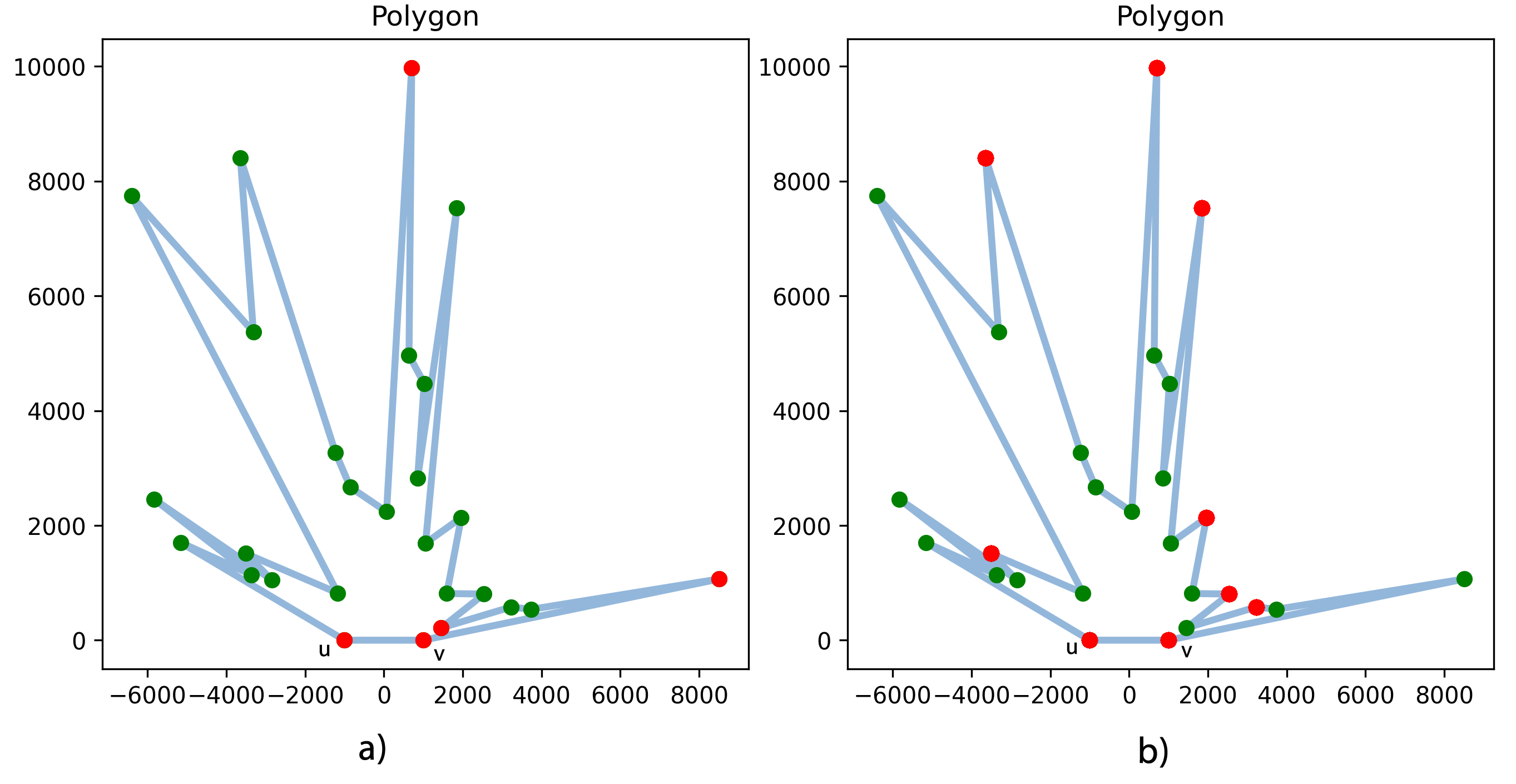}
	\caption{(a) For polygon with $ 27 $ vertices such that $ r = 5 $ size of guard set in Algorithm 1 is $ 3 $ and it runs in 0.75041 s. (b) Size of guard set in Algorithm 2 is $ 9 $ and it runs in 0.323258 s. }
	\label{fig:test2_3}
\end{figure}
Based on outputs (see Figure~\ref{fig:test2_1}, Figure~\ref{fig:test2_2} and Figure~\ref{fig:test2_3}) we understand that for a low value of $ n $ and $ \frac{n}{2} \leq r $ Algorithm 1 is better for guarding a weak visibility polygon with minimum number of guards. because when number of reflex vertices increase, number of diameter of polygon and convex components decrease. So Algorithm 1 can be faster in this situation. Since Algorithm 1 find minimum number of guards, we prefer to use this algorithm for weak visibility polygons with low value $ n $ and $ \frac{n}{2} \approx r $ .

\subsection{Test on large weak visibility polygons}
In the last test, we test arbitrary weak visibility polygons with high value of $ n $ (number of vertices) in Algorithm 1 and Algorithm 2.

\begin{figure}[H]
	\centering
	\includegraphics[scale=0.6]{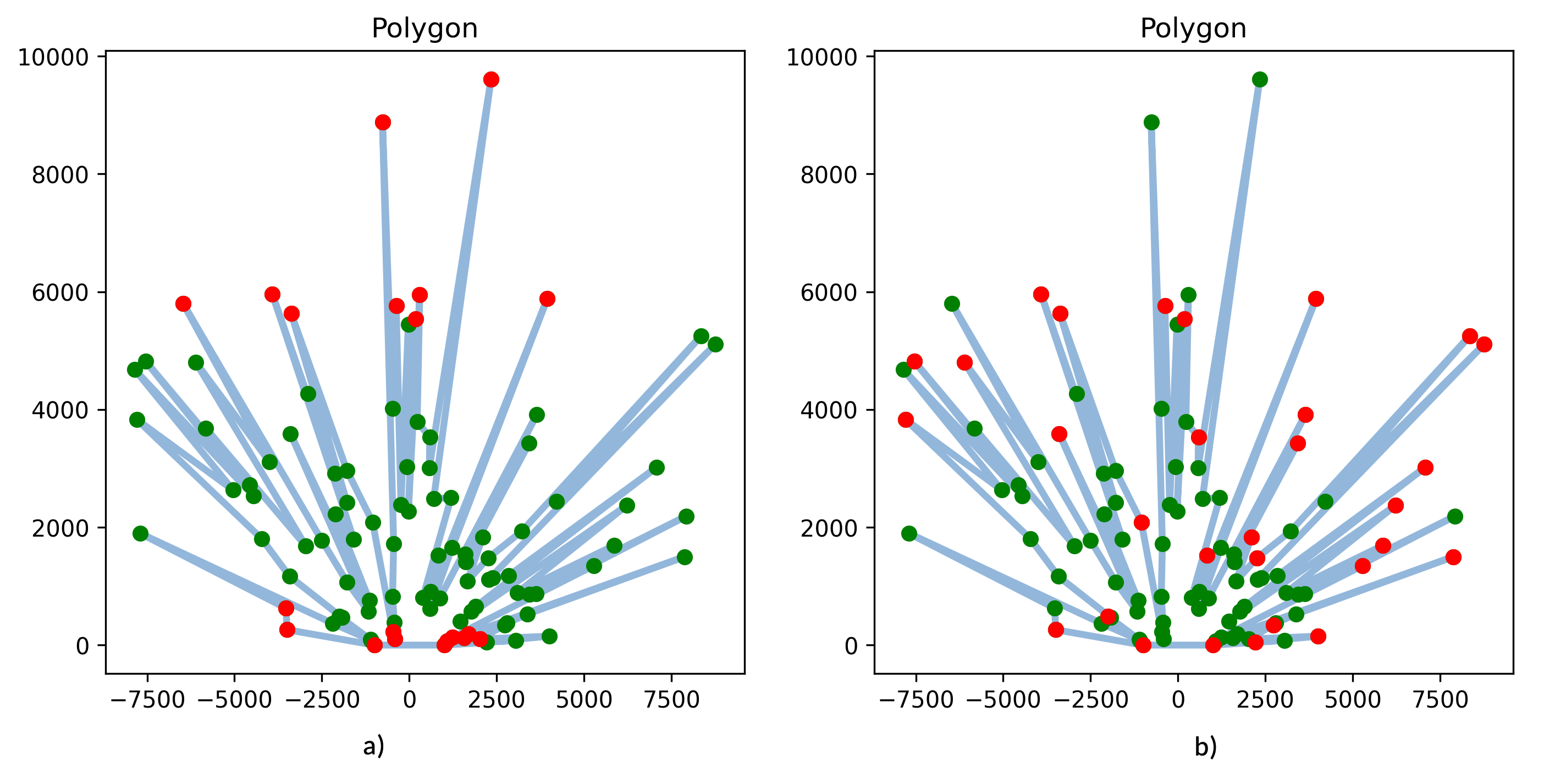}
	\caption{(a) For weak visibility polygon with $ 100 $ vertices, Algorithm 1 uses 20 vertex guards to guard the polygon and it runs in 24.71 s. (b) For weak visibility polygon with $ 100 $ vertices, Algorithm 2 uses 30 vertex guards to guard the polygon and it runs in 15.26 s.}
	\label{fig:test3_1}
\end{figure}

\begin{figure}[H]
	\centering
	\includegraphics[scale=0.5]{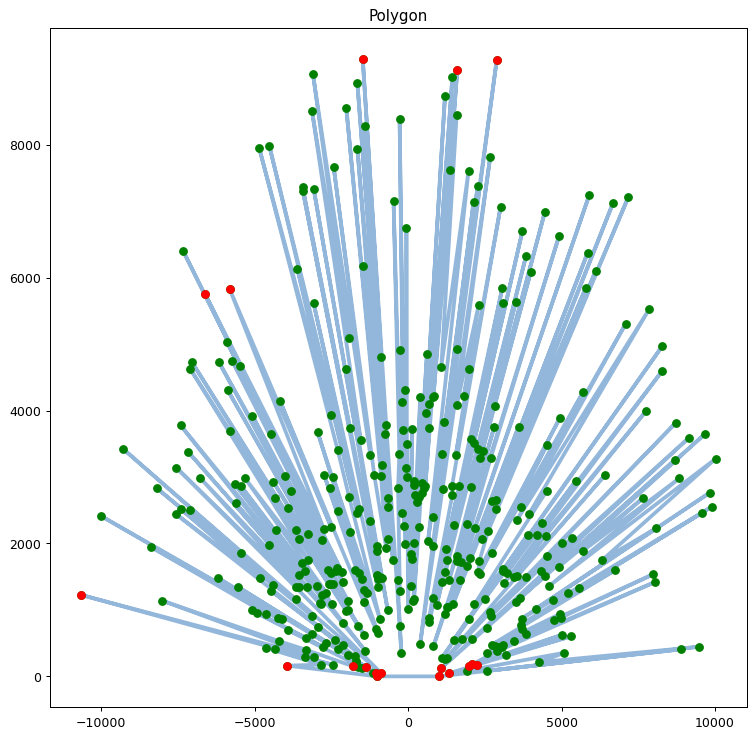}
	\caption{For weak visibility polygon with $ 400 $ vertices, Algorithm 1 uses 18 vertex guards to guard the polygon and it runs in 231.245 s. }
	\label{fig:test3_2_sim}
\end{figure}

\begin{figure}[H]
	\centering
	\includegraphics[scale=0.5]{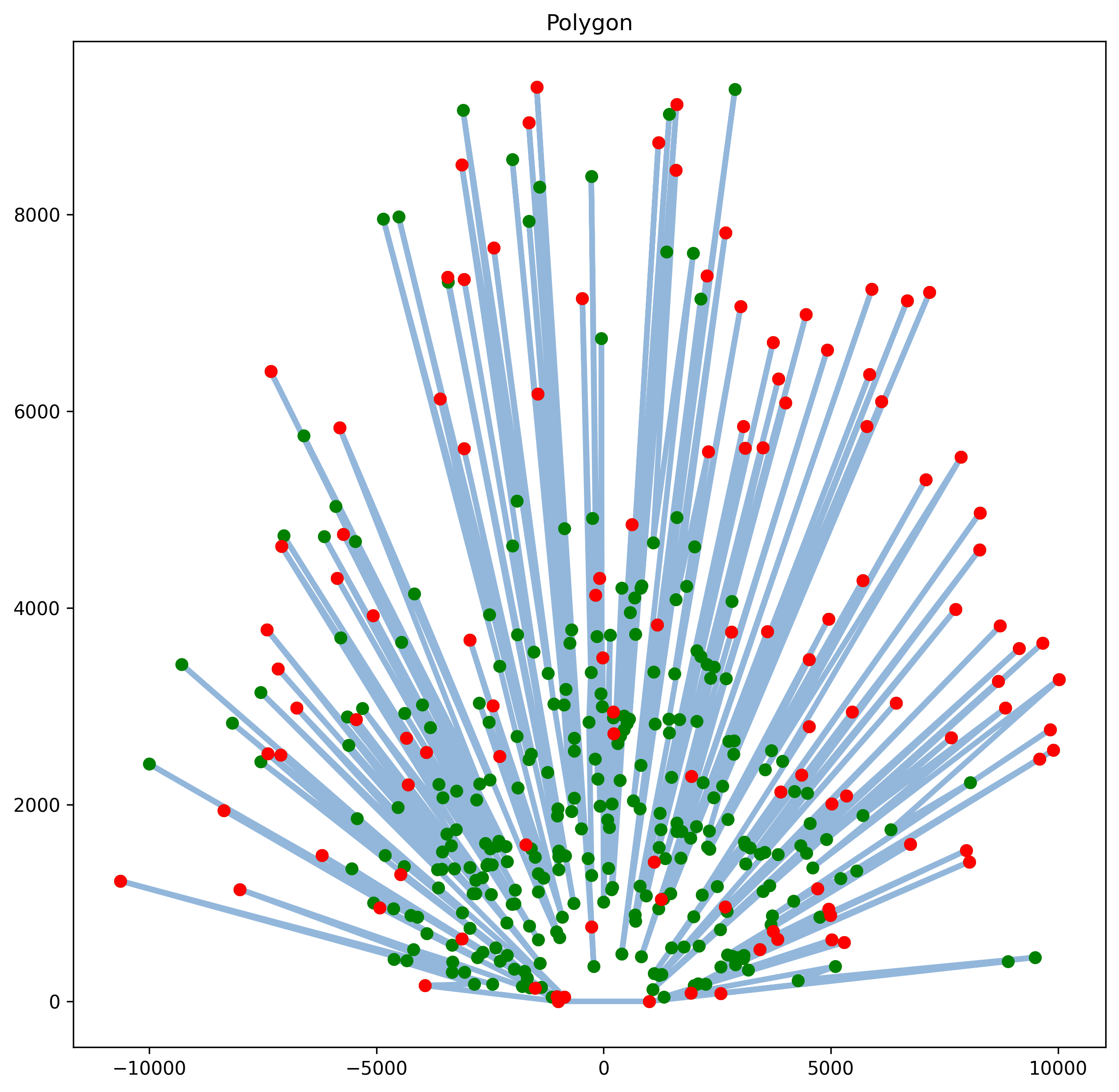}
	\caption{For weak visibility polygon with $ 400 $ vertices, Algorithm 2 uses 115 vertex guards to guard the polygon and it runs in 199.245 s. }
	\label{fig:test3_2_weak}
\end{figure}

Based on outputs (see Figure~\ref{fig:test3_1}, Figure~\ref{fig:test3_2_sim} and Figure~\ref{fig:test3_2_weak}) we understand that for a high value of $ n $ Algorithm 1 is better for guarding a weak visibility polygon with minimum number of guards. because Algorithm 1 use less guards than Algorithm 2 to guard a weak visibility polygon $ P $.

\section{ Test on simple polygons }
In section 2 we found out
algorithm 1 assign less number of guards for weak visibility polygons that algorithm 2. Now we need to test on arbitrary simple polygon.

\subsection{Algorithm for create arbitrary simple polygon }
\begin{figure}[H]
	\centering
	\includegraphics[scale=0.7]{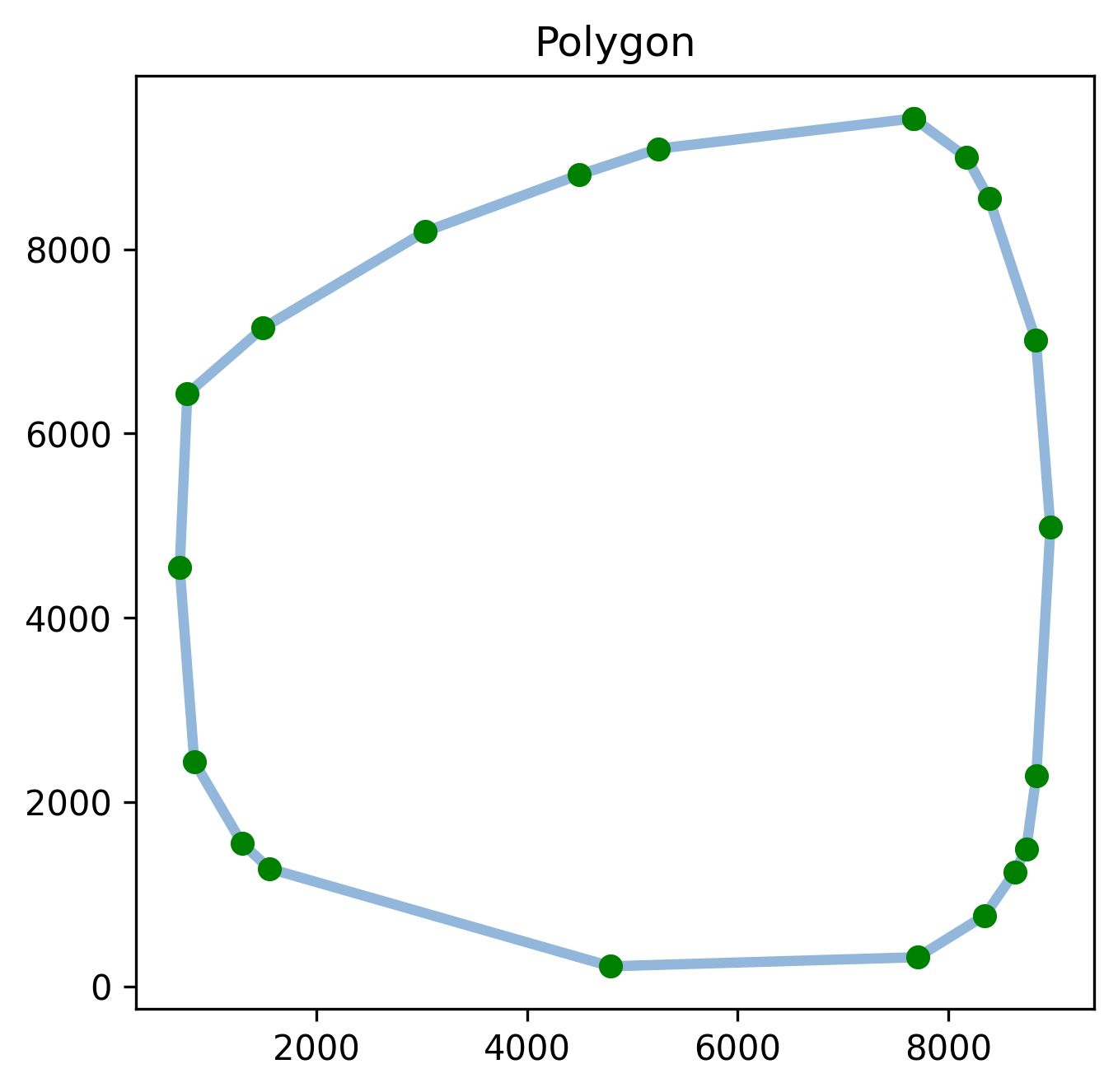}
	\caption{Generate arbitrary convex polygon}
	\label{fig:step1}
\end{figure}

In this section we provide an algorithm that generates simple polygons with custom number of reflex vertices. the sequence of that is like :

1. Generate a simple convex polygon $ P $ with $ n $ number of vertices. (see Figure \ref{fig:step1})

\begin{figure}[H]
	\centering
	\includegraphics[scale=0.7]{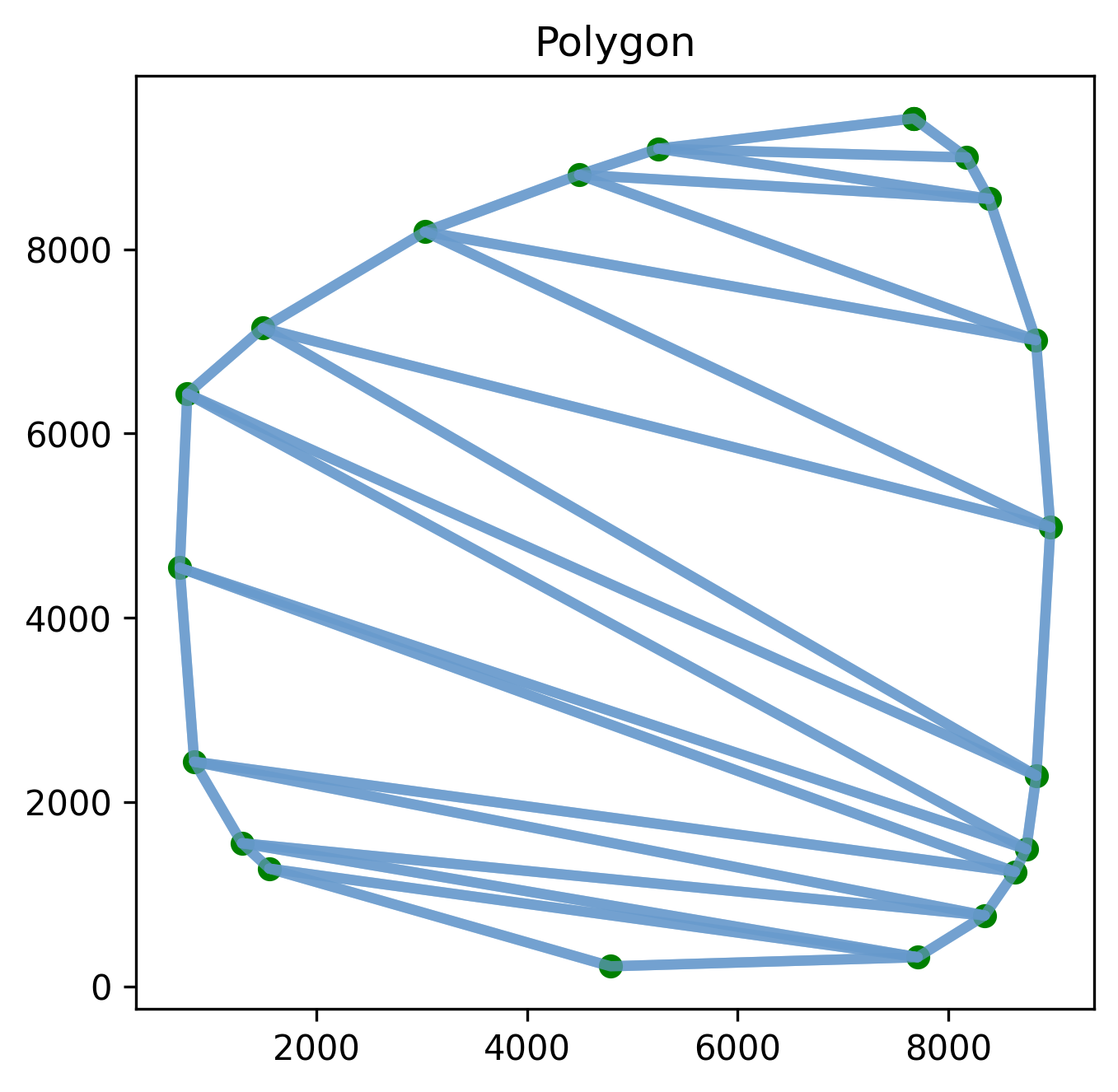}
	\caption{Example of triangulate convex simple polygon $ P $.}
	\label{fig:step2}
\end{figure}

2. Triangulate $ P $ such that every triangle has a joint edge with boundary of $ p $.(see Figure~\ref{fig:step2})

\begin{figure}[H]
	\centering
	\includegraphics[scale=0.7]{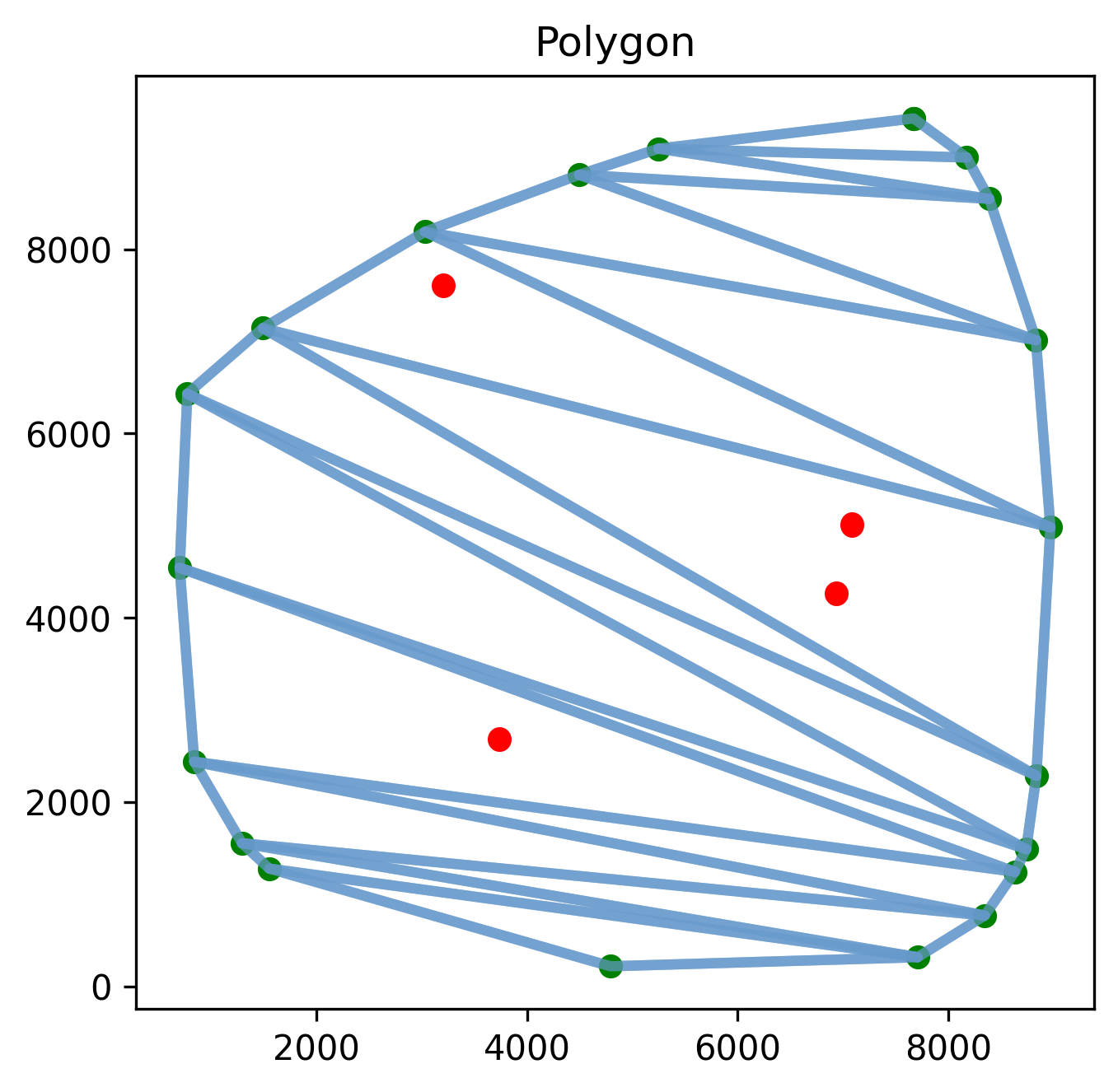}
	\caption{Generate random points in convex polygon}
	\label{fig:step3}
\end{figure}

3. Randomly add points as number of reflex vertices in $ P $. (see Figure \ref{fig:step3})

\begin{figure}[H]
	\centering
	\includegraphics[scale=0.7]{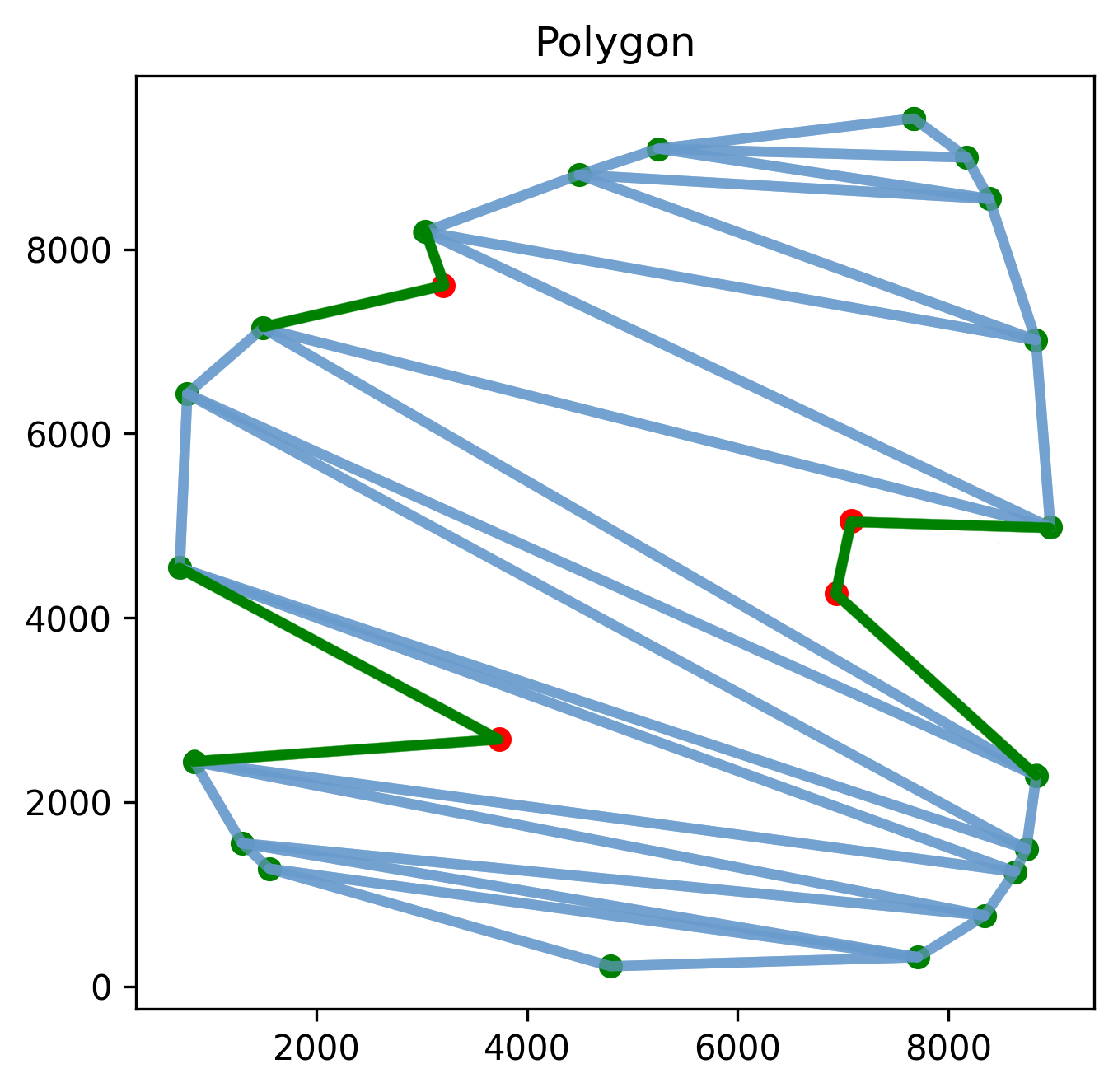}
	\caption{Connect random points to convex polygon and create arbitrary simple polygon with custom number of reflex vertices.}
	\label{fig:step4}
\end{figure}

4. Any such added point $ X $ will be inside a triangle, say $ abc $. Connect $ X $ to $ b $ and $ c $, where $ b $ and $ c $ are two consecutive points on the boundary of $ P $.  If more that one point lies inside $ abc $, connect them by a simple path and then connect the endpoints of the path to $ b $ and $ c $. (see Figure \ref{fig:step4})

This method keeps our polygons simple and create custom number of reflex vertices on that.

\subsection{New Algorithm for guarding}
In this section we want to established an algorithm that is suitable for polygons with number of high value of $ n $.
Conclusions of last tests
observed that if the number $ r $ of reflex vertices is very small
compared to $ n $, then the size of the optimal guard set is $ r $ or close to
$ r $. This means that the number of edges $ E $ in the visibility graph of
such simple polygons is $ O(n^{2}) $. So we choose a small number ($ \log \log n $) as
an upper bound for $ r $ so that $ r $ and optimal are close. 

In this algorithm if the number of reflex vertices is within a very very small fraction (say, $ r-c \leq \log \log n $) of the total number vertices $ n $, where $ c $ is a very small constant, then we place guards at all reflex vertices for guarding a simple polygon $ P $, Otherwise, we place guards using the method of Algorithm 1.

\subsection{Test our new Algorithm}

In this test we use simple polygons as we describe how we generated. We start from polygons with low value of reflex vertices $ r $ then gradually increase $ r $ and distribute reflex vertices around the polygon so the number of edges $ E $ in visibility graph of the polygon gradually
reduces from $ O(n^{2}) $ to $ O(n) $. 
\begin{figure}[H]
	\centering
	\includegraphics[scale=0.4]{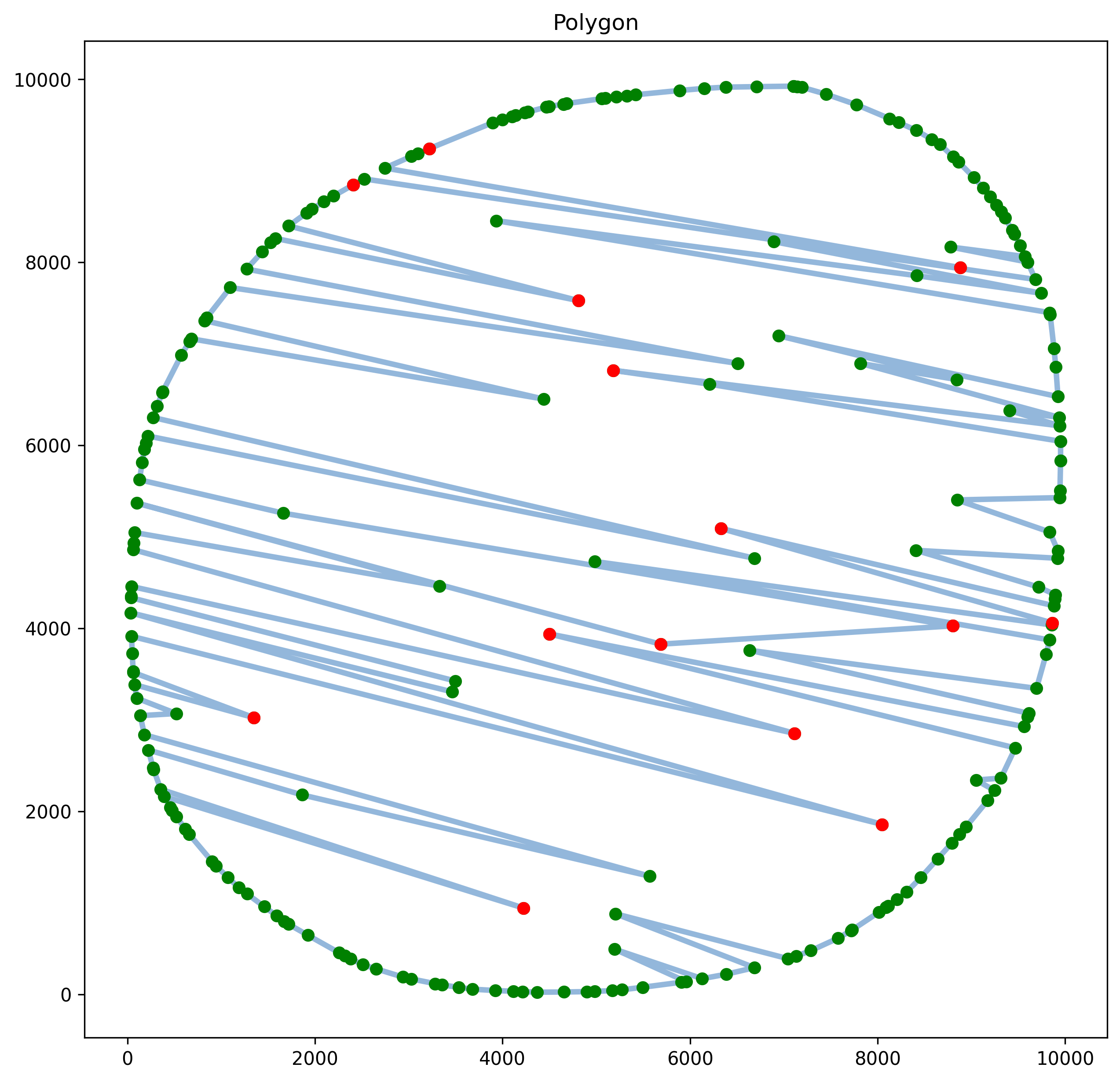}
	\caption{For polygon with $ 219 $ vertices such that $ r = 34 $ guard set in our new algorithm use $ 14 $ guards and it runs in 670.436 s. }
	\label{fig:test4_1}
\end{figure}
\begin{figure}[H]
	\centering
	\includegraphics[scale=0.4]{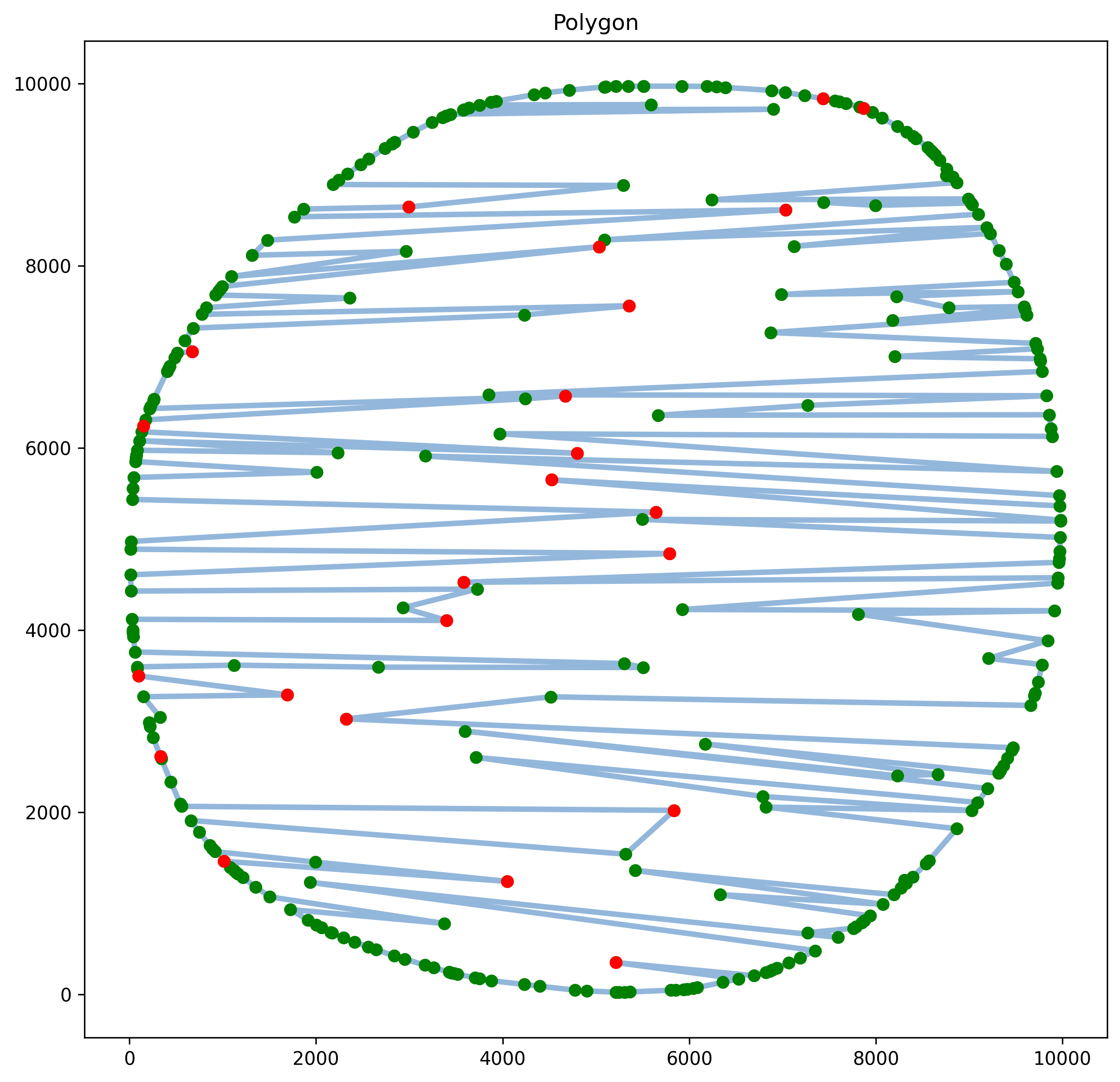}
	\caption{For polygon with $ 300 $ vertices such that $ r = 65 $ guard set in our new algorithm use $ 23 $ guards and it runs in 520.861 s. }
	\label{fig:test4_2}
\end{figure}

The outputs of test suggest that (see Figure~\ref{fig:test4_1} and Figure~\ref{fig:test4_2}) even if $ E $ reduces, the chosen guard  set
remains close to optimal and the algorithm assigns no more than twice the optimal number of
guards.

\section{Conclusion}

In this paper we established that the Ghosh conjecture in 1986 was that through his
approximation algorithm is O(log n) times optimal theoretically, it
will perform much better in practice. All the experimental data that are provided in the entire paper shows that even for complex simple polygons, the chosen guard set by the algorithm is very close to optimal. Therefore, Ghosh's algorithm performs like a constant approximation algorithm in practice.

\subparagraph*{Acknowledgments.} The work is inspired and guided by Prof Subir Kumar Ghosh.

\bibliographystyle{plain}
\bibliography{references}  






\end{document}